\documentclass[12pt]{article}

\usepackage{jheppub}
\usepackage{mathtools,amssymb,amsthm}
\usepackage[utf8]{inputenc}
\usepackage{enumitem}
\usepackage{xcolor}
\usepackage{graphicx}
\usepackage{multirow}
\usepackage[normalem]{ulem}
\usepackage{longtable}
\allowdisplaybreaks
\usepackage{cancel}

\newcommand{\be}{\begin{equation}}
\newcommand{\ee}{\end{equation}}
\newcommand{\bi}{\begin{itemize}}
\def \bea {\begin{eqnarray}}
\def \eea {\end{eqnarray}}

\def\ba#1\ea{\begin{align}#1\end{align}}
\def\bad#1\ead{\begin{aligned}#1\end{aligned}}
\def\bg#1\eg{\begin{gather}#1\end{gather}}
\def\bm#1\em{\begin{multline}#1\end{multline}}
\def\bmd#1\emd{\begin{multlined}#1\end{multlined}}

\newcommand{\ignore}[1]{}

\definecolor{darkgreen}{RGB}{50,150,0}

\def \bal#1\eal  {\begin{align} #1 \end{align}}
\def \bga#1\ega  {\begin{gather} #1 \end{gather}}
\def\({\left(}
\def\){\right)}
\def\[{\left[}
\def\]{\right]}
\def\<{\left\langle}
\def\>{\right\rangle}

\newcommand{\eim}{\end{itemize}}
\newcommand{\beq} {\begin{equation}}
\newcommand{\eeq} {\end{equation}}
\newcommand{\bc}{\begin{center}}
\newcommand{\ec}{\end{center}}

\begin{document}

\title{Casimir Energy Stabilization of Standard Model Landscape in Dark Dimension}
\author[a]{Chuanxin Cui,}
\author[b]{and Sirui Ning}
\affiliation[a]{Department of Physics and Astronomy, Northwestern University, Evanston, Illinois 60208, USA}
\emailAdd{chuanxincui2028@u.northwestern.edu, sirui.ning@physics.ox.ac.uk}
\affiliation[b]{The Rudolf Peierls Centre for Theoretical Physics, University of Oxford,\\Oxford OX1 3PU, UK}

\abstract{In this paper we present a realization of the dark dimension. We consider the 5D Standard Model coupling to gravity with one dimension compactified on an orbifold, which is seen as the dark dimension of size R. We stabilize the radion by Casimir effect wrapping around the compact dimension and recover the neutrino mass and 4D cosmological constant with the observed value. Orbifold can lead to a natural resolution of chirality problem in 5D at low energy, which we briefly discussed in the paper. Although we found that the radion mass is too light to survive under solar system tests of GR, several screening mechanisms might give us a solution, for example, the Chameleon mechanism. 
}

\maketitle


\section{Introduction}

The Standard Model (SM) is perhaps the most successful theory for describing our real world. It offers a unified explanation of electromagnetism, weak interaction, and strong interaction. The bosonic segment of the model features essential components like the Higgs boson and photons, while the fermionic portion includes leptons, quarks, and neutrinos. In terms of degrees of freedom, the bosonic section has 3, which are 1 Higgs Boson and 2 photons, and Majorana neutrinos have 6 degrees of freedom, or 12 degrees of freedom for Dirac fermions. Another critical parameter for understanding our universe is the cosmological constant. This naturally leads to a fundamental question: Are these parameters interrelated, or are they independent of each other? This question delves into the concept of naturalness.\cite{Shaposhnikov:2007nj,Nielsen:2012pu,Masina:2012tz,Degrassi:2012ry,Peccei:1977ur, Shifman:1979if, Kim:2008hd, tHooft:1979rat,Yang:2018iki}

The value of the cosmological constant is intrinsically linked to the magnitude of dark energy. When we have a small cosmological constant, it implies a correspondingly small amount of dark energy. This hints at the possibility that we inhabit a region that borders the quantum gravity landscape, as suggested by the distance conjecture \cite{Ooguri:2006in}. According to the distance conjecture, there should be an infinite light tower of light states, which, in turn, introduces an additional mesoscopic scale. This scale is often referred to as the "dark dimension." \cite{Montero:2022prj}.

In \cite{Arkani-Hamed:2007ryu}, the authors compactify $4D$ Einstein gravity with a cosmological constant on a circle. This results in a $3D$ effective field theory that explains how the degrees of freedom in the Standard Model interact with gravity. The size of the extra dimension is stabilized through a combination of the bosonic Casimir energy, fermionic Casimir energy, and the cosmological constant. An essential aspect of this stabilization involves violating the null energy condition by incorporating the Casimir energy, a crucial step in securing the stability of the extra dimension's moduli. This approach yields a range of distinct vacua, and by finely tuning the parameters, it becomes possible to create an effective field theory that mirrors the observational results of our real world. In this paper, we extend their methodology to compactify $5D$ Einstein gravity on a circle, aiming to provide a concrete construction of the dark dimension. 
It's worth noting that a recent paper \cite{Branchina:2023ogv} also explored the $5D$ dark dimension. In their study, they treated the $5D$ theory as an Effective Field Theory (EFT) and calculated the $4D$ vacuum energy with $5D$ loops contribution, where UV-sensitive terms can arise. In contrast, our approach involves a conventional Casimir energy calculation, as in \cite{Arkani-Hamed:2007ryu}.

It will be interesting to compare our model with other proposals claiming large extra dimensions\cite{Arkani-Hamed:1998wuz,Antoniadis:1998ig, Randall:1998uk, Randall:1999ee}, which focused on solving the EW hierarchy. In \cite{Arkani-Hamed:1998wuz}, the Standard Model degrees of freedom are localised on a 3-brane by type I or type II string. It is focused on how to get a small neutrino mass from a $5D$ space with brane, where right-handed neutrinos propagate at bulk and couples to Higgs field on localized brane. Interaction is thus small since the volume of the brane is much smaller than the volume of the bulk, therefore the neutrino mass can be naturally small. In \cite{Randall:1998uk, Randall:1999ee}, the authors have put forth an alternative mechanism for dealing with extra dimensions. They achieve the stabilization of moduli through a combination of SUSY breaking and the gaugino condensation mechanism. The degrees of freedom in the Standard Model still remain localized on the D-brane. Their approach offers a new way to tackle the hierarchy problem, which stems from the non-factorizable metric. The source of this hierarchy is tied to an exponential function dependent on the size of the extra dimension. The overarching concept of our model can be traced back to the initial one, in which fermions propagating in the bulk and the rest particles in the Standard Model is localized on a 3-brane. This stability is derived from the Casimir energy arising from both bosonic and fermionic degrees of freedom. 

The paper is structured as follows:
In Section 2, we kick off by deriving the effective Lagrangian when transitioning from a $5D$ framework to a $4D$ one, introducing the radion field. We then blend the $5D$ cosmological constant with vacuum Casimir energy to obtain a $4D$ effective cosmological constant, which essentially acts as the effective potential for the radion field. In section 3, our primary focus revolves around stabilizing the radion moduli. This is achieved through the contribution of Casimir energy originating from various sources, including the  $U(1)$ gauge field, graviton, the radion itself, and the bulk fermions- neutrinos. We explore the effects of altering the $5D$ cosmological constant and demonstrate how it leads to the emergence of the correct de Sitter (dS) vacuum configuration for our universe. Additionally, we contemplate an alternative approach by introducing a rolling scalar field that replaces the $5D$ cosmological constant, and find that it still allows for the construction of a dS vacuum. Section 4 shows how our model can be connected with the observed $4D$ SM and tackles two crucial challenges. Firstly, we discuss concerns regarding the potentially too-light mass of our radion field, which could potentially impact gravity. Secondly, we briefly address the chirality problem.
Finally, in Section 5, we wrap up our paper and outline potential avenues for further research and exploration.

\section{Radion vacuum stabilization}
\label{sec1}

We start with the metric on 5-dimensional ($5D$) spacetime in Einstein frame and consider the circle compactification from $5D$ to 4$D$, where we take $x^5=\phi$ to be the compact dimension. The metric is given as:
\begin{equation}
    ds_5^2=\frac{r}{R}ds_4^2+R^2d\phi^2,
    \label{5dm}
\end{equation}
where  $R$ is the radion field, $\phi\in [0, 2\pi)$ and r is an arbitrary positive scale at the moment, which will be fixed to be proportional to the expectation value of R later. 
In writing the metric ansatz in Eq.~(2.1) we introduced a constant factor $r$ multiplying the non-compact line element. 
This $r$ should be understood as a \emph{fiducial four-dimensional length/volume normalization} that keeps track of the (infinite) $4d$ volume when reducing the action.
In Minkowski compactifications, $r$ can be removed by a trivial rescaling of the $4d$ coordinates and therefore drops out of all physical quantities.

When the $5d$ vacuum energy $\Lambda_5$ is present, the physically relevant quantity is the \emph{$4d$ action density} (or equivalently the effective potential per unit physical $4d$ volume) evaluated in the $4d$ Einstein frame.
When taking the on-shell condition, the $4d$ curvature scale (or the resulting $4d$ cosmological constant) is fixed self-consistently by the $4d$ Einstein equation at the extremum, and all predictions can be expressed in terms of $\Lambda_5$, the stabilized radius $R_\star$, and the resulting $\Lambda_4$.
In particular, once one works with the $4d$ action density, any dependence on the arbitrary normalization $r$ disappears.
The Einstein-Hilbert action governing the dynamics in $5D$ theory is 
\begin{equation}
\begin{aligned}
    L=&\int dx^{4}d\phi\sqrt{-g_5}\left(\frac{1}{2}M_{5}^{3}R_{5}-\Lambda_{5}\right),\\ 
    \end{aligned}
    \label{5lag}
\end{equation}
where $M_5$ is the $5D$ Planck mass, $R_5$ is the Ricci scalar in $5D$ spacetime and $\Lambda_5$ is the $5D$ cosmological constant. 

At distances larger than $R$, we can get an effective 4$D$ action by dimensional reduction. With the parametrization in Equ.(\ref{5dm}), the relationship between Ricci scalar in $5D$ and $4D$ goes like,
\begin{equation}
\begin{aligned}
    R_{5}&=\frac{R}{r} R_{4}-\frac{3R}{2r}\Big{(}\frac{\partial R}{R}\Big{)}^2 -2\frac{1}{R}\partial^2 R, \\
\end{aligned}
\label{5Ric}
\end{equation}

Integrating over $\phi$ and replacing the Ricci scalar and the metric into $4D$ version by Equ.($\ref{5dm}$) and Equ.(\ref{5Ric}), the $4D$ effective action reads (ignoring total derivative term):

\begin{equation}
\begin{aligned}
    L&=\int dx^4 2\pi  \frac{ r^2}{R} \sqrt{-g_4}\left(\frac{1}{2}M_5^3 (\frac{R}{r} R_4-\frac{3R}{2r}\Big{(}\frac{\partial R}{R}\Big{)}^2)-\Lambda_5\right)\\
    &=\int dx^4 \sqrt{-g_4} \left(\frac{1}{2}M_P^2  R_4-  \frac{3}{4}M_P^2 \Big{(} \frac{\partial R}{R} \Big{)}^2 - \Lambda_4\right),\\  
    \end{aligned}
\end{equation}
with $M_p$ ($=1/8 \pi G_N$) and $\Lambda_4$ being the Planck mass and cosmological constant in $4D$, defined as:
\begin{equation}
  M_P^2 = 2\pi r M_5^3,\quad \Lambda_4 = \frac{2\pi r^2}{R}\Lambda_5,
    \label{45rel}
\end{equation}
To obtain a canonically normalized kinetic term for radion field, we define:
\begin{equation}
    R=e^{\sqrt{\frac{2}{3}}\frac{\chi}{M_{P}}},
    \label{canR}
\end{equation}
then the action becomes:
\begin{equation}
    L=\int dx^4 \sqrt{-g_4}\left(\frac{1}{2}M_P^2 R_4 -\frac{1}{2}\partial_u\chi\partial^u\chi-\Lambda_4\right).
\end{equation}


The existence of $4D$ cosmological constant would make the classical potential for radion runaway and then decompactify the circle. In order to stabilize the compactified dimension, we consider the Casimir energy for particles wrapping around the circle, as in Ref~\cite{Arkani-Hamed:2007ryu}. Let's first consider a general case where a massive scalar field is coupled to $5D$ GR with one compacted dimension. The lagrangian reads:
\begin{equation}
\begin{aligned}
    L=&\int dx^{4}d\phi\sqrt{-g_5}\left(\frac{1}{2}M_{5}^{3}R_{5}-\Lambda_{5}+L_{M}\right),\\ 
    \end{aligned}
\end{equation}
where $L_M$ refers to a free massive scalar field:
\begin{equation}
    L_{M}=-\frac{1}{2}\partial_u\phi\partial^u\phi-\frac{1}{2}m^2\phi^2.
\end{equation}

The expectation value of energy momentum tensor takes the form (for detailed calculation, see \ref{appen}):
\begin{equation}
\begin{aligned}
    T_{\mu\nu}&=\left\langle L_{M}g_{\mu\nu}-2\frac{\delta L_{M}}{\delta g^{\mu\nu}}\right\rangle\\
    &=-\left(\rho(R)\eta_{uv}+R\rho'(R)\delta_u^\phi\delta_v^\phi \right),\\
    \end{aligned}
    \label{emt}
\end{equation}
where $\rho(R)$ is the Casimir energy density, which is defined by summing over infinite volume Green functions $G_{\infty}$:
\begin{equation}
    \rho(R)=2\sum_n \frac{\partial G_{\infty}(y_n^2)}{\partial y_n^2}\Big{|}_{y_n=2\pi Rn\hat{\phi}}.
\end{equation}
Note that the above summation over all integers n must exclude $n=0$ case, which just corresponds to usual $R$-independent Casimir contribution to cosmological constant. 

The explicit Casimir energy density formula for a massless field in D=5  reads:
\begin{equation}
    \rho(R)=-\frac{3}{4\pi^2}\zeta(5)\frac{1}{(2\pi R)^5},
\end{equation}
where $\zeta(5)=1.036927..$ is the Riemann zeta function at $z=5$.  For massive field, the Casimir energy density takes the form:
\begin{equation}
    \rho(m,R)=-\sum_{n=1}^{\infty}\frac{2m^5}{(2\pi)^{\frac{5}{2}}}\frac{K_{\frac{5}{2}}(2\pi Rmn)}{(2\pi Rmn)^{\frac{5}{2}}},
\end{equation}
where $K_{\nu}(z)$ is the modified Bessel function of the second kind
\begin{equation}
    K_{\nu}(z)=\frac{1}{2}\int^{\infty}_{0} d \beta \beta^{\nu-1} e^{-\frac{z}{2}(\beta+\frac{1}{\beta})}.
    \label{Besselk}
\end{equation}
Note that above formula of Casimir energy density used periodic boundary conditions. So it only applies to bosonic field. For fermionic field, the energy contribution will differ with an extra minus sign.


Employing the formula of energy momentum tensor calculated in Equ.(\ref{emt}), the $5D$ Einstein equation reads (for $\mu,\nu=1,2,3,4$):
\begin{equation}
    M_5^3(R_{\mu\nu}-\frac{1}{2}Rg_{\mu\nu})+\Lambda_5 g_{\mu\nu}=T_{\mu\nu}=-\rho(R)\eta_{\mu\nu},
\end{equation}
which can be rewritten as:
\begin{equation}
     M_5^3(R_{\mu\nu}-\frac{1}{2}Rg_{\mu\nu})+\Lambda_{5 \rm eff}\eta_{\mu\nu}=0,
\end{equation} 
where we've defined an effective cosmological constant:
\begin{equation}
    \Lambda_{5 \rm eff}=\Lambda_5+\rho(R)
\end{equation}

Following the parametrization in Equ.(\ref{45rel}), the $4D$ effective cosmological constant reads:
\begin{equation}
    \Lambda_{4 \rm eff}=\frac{2\pi r^2}{R}(\Lambda_5+\rho(R))
\end{equation}
which is equivalent to $4D$ effective potential for radion field. Therefore the explicit formula for radion potential goes like:
\begin{equation}
    \begin{aligned}
       V(R)&=r^2 \left( -\frac{3n_B}{64\pi^6R^6}\zeta(5)+\frac{2\pi \Lambda_5 }{R}+\frac{3n_F^1}{64\pi^6R^6}\zeta(5)- \frac{2\pi}{R} \sum_{i=2}n_{F}^i \rho(m_i,R) \right),
    \end{aligned}
    \label{radpoten}
\end{equation}
with $n_B$, $n_F^1$ and $n_F^i$ ($i\geq$ 2) being the physical degrees of freedom (d.o.f.) for massless bosonic field, massless fermionic field and massive fermionic fields with mass $m_i$ respectively. It is worth noting that the contribution from massless field, massive field and cosmological constant shows three different dependence on $R$, which is a necessary condition for the vacuum to be stabilised.

\section{Extracting de-Sitter vacua}
\label{result}

Radion effective potential is closely related to the number of d.o.f. of different fields, as it is shown in Equ.(\ref{radpoten}). In Standard Model (SM), the only massless particles are graviton and photon. For a general $D$ dimensional spacetime, physical d.o.f. of vector (massless $U(1)$ gauge field) and traceless symmetric (graviton) representation of massless fields are given as:
\begin{equation}
    \begin{aligned}
        &\textbf{ $U(1)$ gauge field}: D-2,\\ 
        &\textbf{Graviton}: \frac{(D-1)(D-2)}{2}-1.\\
    \end{aligned}
\end{equation}
We will consider a \emph{massless $U(1)$ gauge field} propagating in the 5d bulk, which contributes to the Casimir energy.
We stress that we do \emph{not} identify this field with the Kaluza--Klein vector coming from the metric (often called the graviphoton), nor do we assume it is the Standard Model photon; it can be a generic bulk/hidden $U(1)$.
Identifying it with the SM photon would require additional model-building ingredients and is not assumed here.
Therefore, $n_B=5+3=8$ at $D=5$. Without the contribution from fermions, the potential becomes (take $\zeta(5) \approx 1$):
\begin{equation}       
    V(R)=r^2 (-\frac{3}{8 \pi^6 R^6} +\frac{2\pi \Lambda_5 }{R}),
\end{equation}
which gives the critical radius when $V$ takes a maximum at $R=R_{c}$:
\begin{equation}
    R_{c}=\left(\frac{9}{8 \pi^7 \Lambda_5}\right)^{\frac{1}{5}}.
\end{equation}

To stabilize the vacuum, we assume the size of $1/R_c$ is small enough to be comparable with the lightest massive particle in Standard Model- neutrinos (which can always be achieved by adjusting the size of $\Lambda_5$ properly). Note that more massive states in SM ($m >$ $1/R$) contribute to Casimir energy at the next leading order. The mass splittings of neutrino spectrum for solar and atmospheric oscillations in normal hierarchy spectrum are \cite{ParticleDataGroup:2022pth}:
\begin{equation}
    \begin{aligned}
        \Delta m_{solar}^2&= \Delta m_{12}^2 \approx(7.53\pm 0.18) \times 10^{-5} eV^2,\\
        \Delta m_{atm}^2 &= \Delta m_{23}^2 \approx (2.437 \pm 0.033)\times 10^{-3}eV^2.
    \end{aligned}
\end{equation}
We then add the contribution of three neutrinos to the Casimir energy. Note that in 5 dimensional spacetime, only Dirac fermions can exist. For an odd $D$ dimensional spacetime, the d.o.f for a massive Dirac fermion is given by:
\begin{equation}
     \begin{aligned}
        &\textbf{Massive Dirac Fermion}: 2^{k+1},\\ 
    \end{aligned}
\end{equation}
where $D=2k+3$. Hence $n_F$=4 at $D=5$. Since we only know the mass splittings for three neutrinos, we take our input masses for neutrinos as follows:
\begin{equation}
    m_1=0, \quad m_2=\Delta m_{12}= 0.008678 eV, \quad m_3= \Delta m_{12}+\Delta m_{23}= 0.058044 eV.
\end{equation}
Therefore, the effective potential for radion in Equ.(\ref{radpoten}) now reads:
\begin{equation}
    \begin{aligned}
       V(R)&=r^2 \left( -\frac{3}{8\pi^6R^6}\zeta(5)+\frac{2\pi \Lambda_5 }{R}+\frac{3}{16\pi^6R^6}\zeta(5) - \frac{8\pi}{R} \rho(m_2,R) - \frac{8\pi}{R} \rho(m_3,R) \right).
    \end{aligned}
    \label{radpoten2}
\end{equation}
The neutrino contribution to the Casimir energy depends on the masses only through the dimensionless combinations $m_i R$.
For $m_i R \ll 1$ the neutrino Casimir term approaches the massless-fermion limit, while for $m_i R \gg 1$ it becomes exponentially suppressed.
As a result, varying the neutrino splittings (including taking $m_1>0$ so that all three neutrinos are massive) does not change the qualitative structure of the effective potential; it shifts the position and depth of the extremum smoothly.
In particular, a dS minimum persists as long as at least one neutrino satisfies
\begin{equation}
m_i R_\star \lesssim \mathcal{O}(1)
\end{equation}
near the stabilized radius $R_\star$, so that the positive massive-fermion Casimir contribution is not exponentially suppressed.
If instead all three neutrinos satisfy $m_i R_\star \gg 1$, their contribution is negligible and the dS minimum may disappear.

\begin{figure}[t]
  \begin{center}
   \includegraphics[width=10cm]{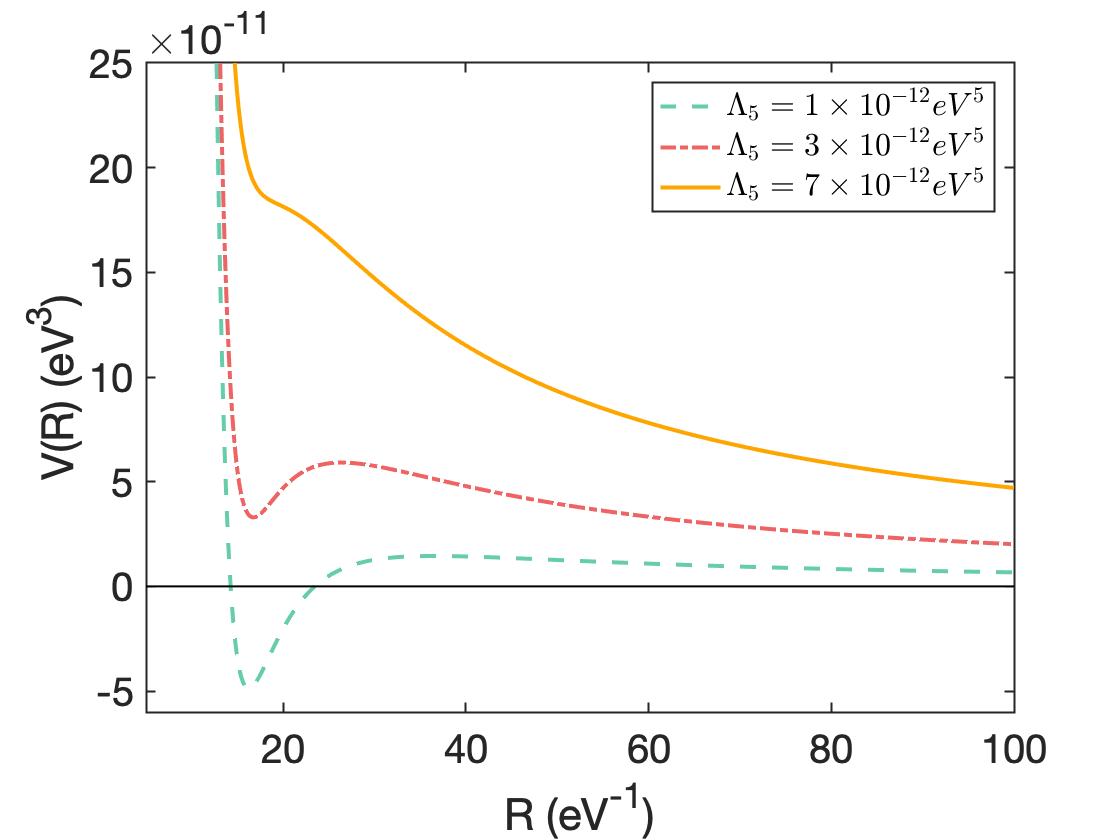}
  \end{center}   
\vspace*{-0.5cm}
\caption{Effective potential for radion field in the case of three different value of $\Lambda_5$: a) $\Lambda_5=1\times 10^{-12}eV^5$ with an anti de-Sitter vacuum at $R=16.3$ $eV^{-1}=3.2156\mu m$. b) $\Lambda_5=3\times 10^{-12}eV^5$, which produced a de-Sitter vacuum at $R=16.8$ $eV^{-1}=3.3146\mu m$. c) $\Lambda_5=7\times 10^{-12}eV^5$ with runaway potential. In the plot we've fixed $r^2=107 eV^{-2}$ so that we can reproduce the correct value for $\Lambda_4$ in de-Sitter case at critical $R$: $\Lambda_4=V(R_c)\approx 3.3 \times 10^{-11} eV^4$.} 
\label{potential-fig}
\end{figure}

\begin{figure}[t]
  \begin{center}
   \includegraphics[width=10cm]{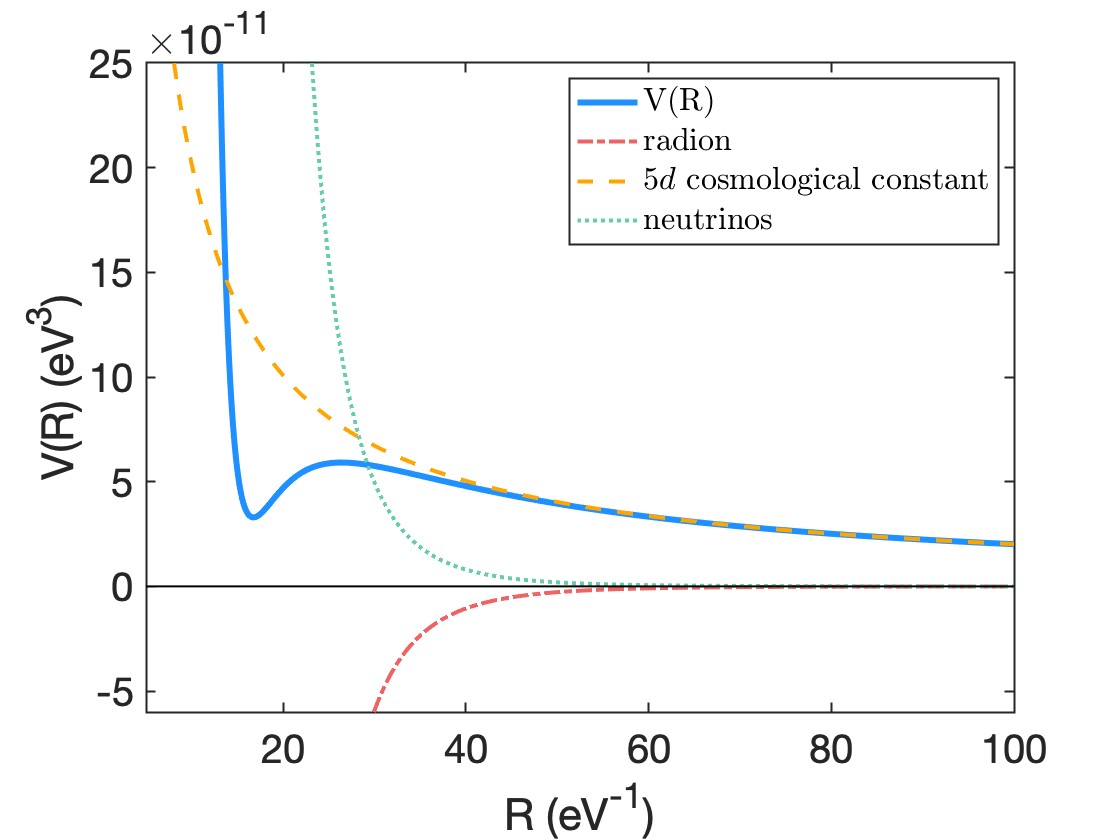}
  \end{center}   
\vspace*{-0.5cm}
\caption{Contributions from different fields are shown. In the plot we took $\Lambda_5=3\times 10^{-12}eV^5$.} 
\label{potential-fig3}
\end{figure}

\begin{figure}[t]
  \begin{center}
   \includegraphics[width=10cm]{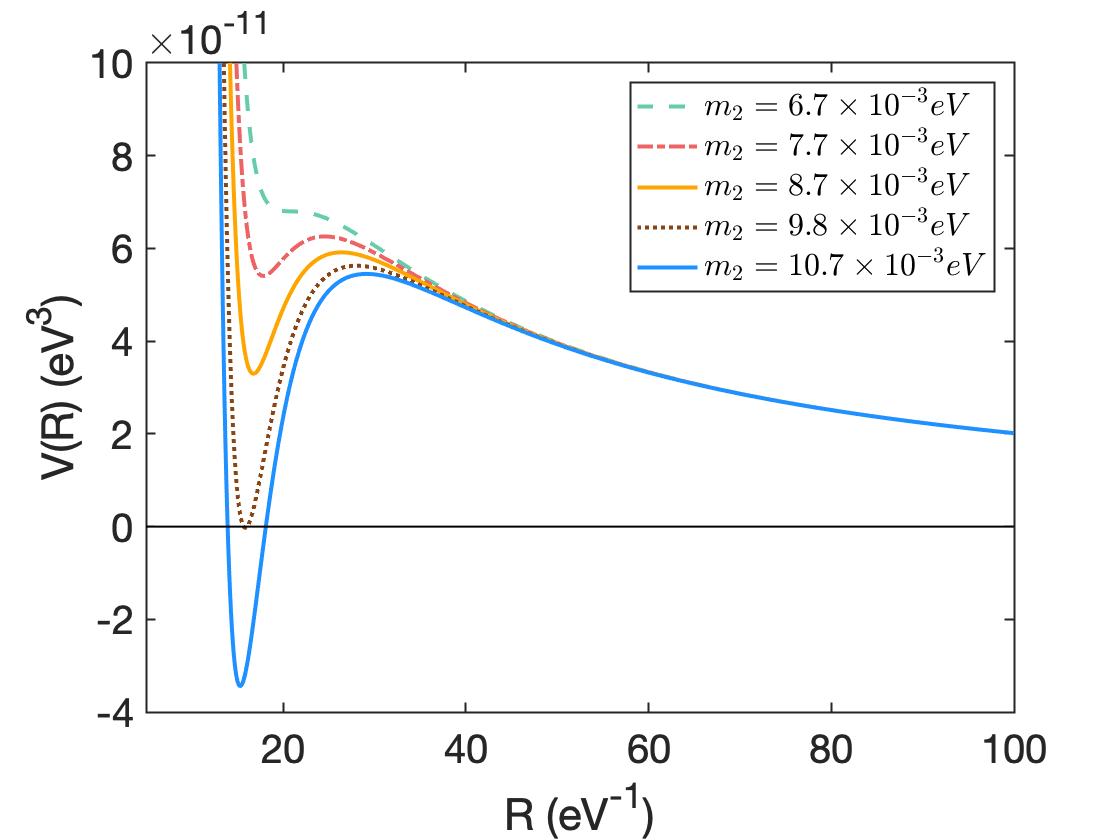}
  \end{center}   
\vspace*{-0.5cm}
\caption{Radion 4D potential in the case of different input masses of Dirac neutrinos: $m_2$ =6.7, 7.7, 8.7, 9.7, 10.7 $\times 10^{-3}eV$, while $m_3$ = 5.8$\times 10^{-2}eV$, $\Lambda_5=3\times 10^{-12}$ and $r^2=107 eV^{-2}$.} 
\label{potential-fig2}
\end{figure}

Fig.(\ref{potential-fig}) shows the $R$ dependence of $V(R)$ in the cases of three different $\Lambda_5$ value, where we've fixed $r^2=107 eV^{-2}$ and take three neutrino masses in normal hierarchy with $m_1=0$. At $R=R_{c}= 16.8  eV^{-1}\approx 3.3146 \mu m$ with $\Lambda_5 =3\times 10^{-12}eV$, we can reproduce a de-Sitter vacua with the correct 4D cosmological constant $\Lambda_4=V(R_c)\approx 3.3 \times 10^{-11} eV^4$. The contribution from different fields goes like follows: At small $R$, neutrino Casimir energy ($\sim |\rho (m,R)|/R$) takes dominant contribution on $V(R)$. As $R$ gets larger, potential starts to be dragged down by negative contribution from radion Casimir energy ($\sim-1/R^6$). Immediately following, potential is pulled back by positive contribution from $5d$ cosmological constant ($\sim 1/R$) and it keeps the dominant contribution at large $R$. We show the contribution of each terms in Fig.\ref{potential-fig3}.

In Fig.(\ref{potential-fig2}), we explored how potential depends on neutrino masses, where we varied $\Delta m_{solar}^2$ and fixed $\Delta m_{atm}^2$. We take $\Lambda_5=3\times 10^{-12}$ and $r^2=107 eV^{-2}$ in the figure. As a result, we can get an AdS vacuum if the neutrino mass $m_2$ is heavier than $9.8\times 10^{-3}eV$, a dS vacuum when $9.8\times 10^{-3}eV > m_2 > 6.7 \times 10^{-3}eV$ and no stationary vacuum if $m_2 < 6.7 \times 10^{-3}eV$. When we take $m_2=\Delta m_{solar}=$ current experiment value, we could again recover the correct value of $4D$ cosmological constant.

At the end of this section, we briefly discuss an alternative point of view. In our previous discussion, we assumed a bare cosmological constant in $5D$ and have found its $4D$ effective formula. We emphasize that fully controlled top-down de Sitter constructions in five dimensions are scarce and highly constrained in the classes of setups relevant to the dark-dimension scenario.
Nevertheless, de Sitter critical points in $5d$ have been discussed in the recent literature; see e.g.\ Ref.~\cite{BentoMontero2025} for an explicit $dS_5$ maximum obtained from Casimir effects in an M-theory compactification.
Our discussion below concerns a different (effective-field-theory) mechanism for generating a $4d$ dS minimum for the radion.
 Therefore, we might consider a rolling scalar field to replace the $5D$ cosmological constant.

The potential of such rolling field takes the form: $V(\phi)= \Lambda_{\phi} e^{- \alpha \phi}$, where $\phi$ is the rolling field,  $\Lambda_{\phi}$ is the overall coupling constant and $\alpha$ is an arbitrary constant. For concreteness we took $\alpha=1$ in this illustrative parametrization; none of the qualitative points in this paragraph relies on this precise value.
One may instead choose $\alpha=\mathcal{O}(1)$ and, if desired, values compatible with swampland-motivated expectations for exponential potentials (see e.g.\ \cite{Rudelius2021}).
 The role of this sector here is purely phenomenological: it provides a simple way to model a slowly varying effective 5d vacuum energy.A realistic completion should also address the dynamics and couplings of $\phi$.
If $\phi$ remains light and couples (even weakly) to visible matter, it can mediate fifth forces and may lead to late-time evolution that could end the quasi-dS epoch.
Therefore an additional stabilization/screening mechanism, or sufficiently suppressed couplings to the Standard Model sector, would be required.
We leave a detailed analysis of such model-building options to future work.

This new scalar field takes two new contribution to our effective potential for radion: (i) a new bosonic degree of freedom, (ii) rolling potential $\Lambda_{\phi} e^{- \alpha \<\phi\>}$ where $\<\phi\>$ is the expectation value. Since $\Lambda_{\phi}$ would be much smaller than $1/R_c$ (as can be checked in our following result), we could safely assume that such scalar field is massless. Hence the effective potential now reads:

\begin{equation}
    \begin{aligned}
       V(R)&=r^2 \left( -\frac{27}{64\pi^6R^6}\zeta(5)+\frac{2\pi \Lambda_{\phi} e^{-\alpha\<\phi\>} }{R}+\frac{3}{16\pi^6R^6}\zeta(5) - \frac{8\pi}{R} \rho(m_2,R) - \frac{8\pi}{R} \rho(m_3,R) \right).
    \end{aligned}
    \label{radpoten_ro}
\end{equation}

\begin{figure}[t]
  \begin{center}
   \includegraphics[width=10cm]{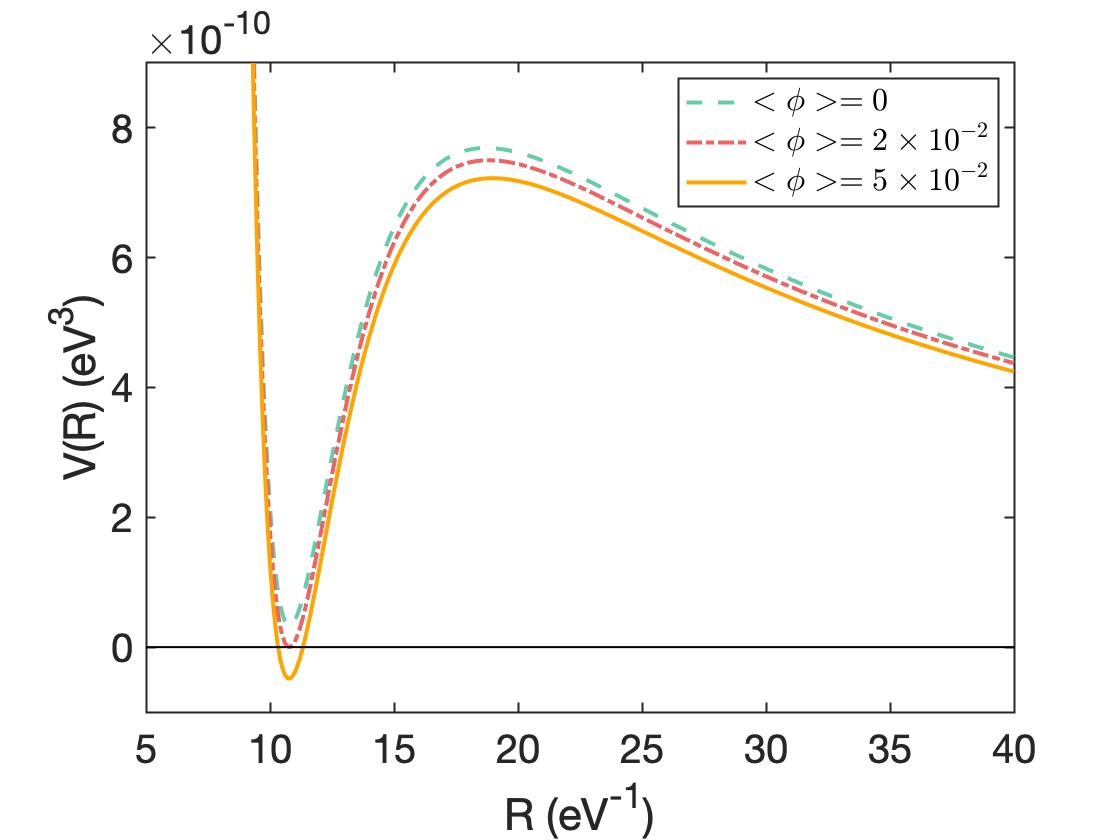}
  \end{center}   
\vspace*{-0.5cm}
\caption{Radion field effective potential under three different value of $\< \phi \>$: 0, $2\times 10^{-2}$, $5\times 10^{-2}$. In the plot we set $r^2=107 eV^{-2}$ and neutrino masses in normal hierarchy with $m_1=0$ as before, and take $\Lambda_{\phi}=2.68\times 10^{-11} eV^5$. } 
\label{potential_rol}
\end{figure}

In Fig.(\ref{potential_rol}) we show our result under different $\< \phi \>$ values. The coupling $\Lambda_{\phi}$ has been set to $2.68\times 10^{-11} eV^5$ in order to reproduce the correct 4D cosmological constant as $\< \phi \>=0$. We see that such rolling field is able to produce ds, flat and AdS vacuum respectively as $\< \phi \>$ increases its value.

\section{Connecting 5D to real world}
\label{chirality}

To move our model closer to the real world, many issues might arise either by constraint from gravitational observation or the speciality of 5D spacetime.

On one side, the mass of our radion field is too light. To obtain the mass of the radion, we need to use the canonical normalized field defined in Equ.(\ref{canR}). The effective potential now reads:
\begin{equation}
    \begin{aligned}
        V(\chi)= & r^2 \Big{(} -\frac{3}{8\pi^6 e^{2\sqrt{6}\frac{\chi}{M_{P}}}}\zeta(5)+\frac{2\pi \Lambda_5 }{e^{\sqrt{\frac{2}{3}}\frac{\chi}{M_P}}}+\frac{3}{16\pi^6 e^{2\sqrt{6}\frac{\chi}{M_{P}}}}\zeta(5)- \frac{8\pi}{e^{\sqrt{\frac{2}{3}}\frac{\chi}{M_P}}} \rho(m_2,e^{\sqrt{\frac{2}{3}}\frac{\chi}{M_P}}) \\
        &- \frac{8\pi}{e^{\sqrt{\frac{2}{3}}\frac{\chi}{M_P}}} \rho(m_3,e^{\sqrt{\frac{2}{3}}\frac{\chi}{M_P}}) \Big{)}.
    \end{aligned}
\end{equation}
Hence the mass of radion can be obtained:
\begin{equation}
    m_R= \sqrt{\frac{\partial^2 V(\chi)}{\partial \chi^2}} \Big{|}_{R=R_c, \Lambda_5 = 3\times 10^{-12} eV, r^2=107 eV^{-2}} =3.07*10^{-33}eV.
\end{equation}
The mass of radion we obtained is below the $4D$ Hubble scale so it might modify our observed gravity. In order to be consistent with solar system tests of GR, some screening mechanisms are required. Several screening mechanisms have been proposed to address the issue of extra fields: Vainshtein mechanism\cite{Vainshtein:1972sx}, Chameleon mechanism \cite{Khoury:2003aq,Khoury:2003rn, Quevedo:1996sv,Quevedo:2002xw,Quevedo:1997uy, Brax:2023qyp}, Symmetron mechanism \cite{Hinterbichler:2010es}, Axion Homeopathy \cite{Burgess:2021qti} and so on. For example, Chameleon mechanism proposed that the effective mass of a scalar field might depend on the environment, i.e. the matter density, due to nontrivial coupling to matter motivated from string theory. Hence the effective mass of light scalar field could be quite large in our solar system region due to high matter density, thus escaped from solar system tests of GR.

Chameleon mechanism requires an $R$-dependent potential for environment matter fields. There is a simple way to obtain this potential just by doing compactification from $5D $ to $4D$. We start with a $5D$ Einstein-Hilbert action with a matter field $\psi$:
\begin{equation}
    S_5=\int d^4 d\phi \sqrt{-g_5} \left[\frac{1}{2}M_5^3 R_5 -\Lambda_5 + \mathcal{L}_M^{\prime} (\psi, g_5) \right]
\end{equation}
The $5D$ metric is given by Equ.(\ref{5dm}). We then integrate over the compacted dimension $\phi$ to obtain $4D$ effective action:
\begin{equation}
     S_4=\int d^4 x  \sqrt{-g_4} \left[\frac{1}{2}M_P^2 R_4 - \frac{3}{4}M_P^2(\frac{\partial R}{R})^2- \Lambda_4 \right] + \int d^4 x
     \sqrt{-\Tilde{g_4}} 2\pi R \left(\mathcal{L}_M (\psi, \Tilde{g_4})+ ...\right) 
     \label{cha1}
\end{equation}
where $\Tilde{g_4}=\frac{r}{R}g_4$ and "..." includes matter component in $\phi$-dimension, i.e. KK modes. Next we redefine the matter field $\psi$ as $\psi^{\prime} \equiv \sqrt{2\pi R} \psi$. Then matter part in Equ.(\ref{cha1}) will be cast into the form:
\begin{equation}
    S_{m}=\int d^4 x
     \sqrt{-\Tilde{g_4}} \left(\mathcal{L}_M^{kin} (\psi^{\prime}, \Tilde{g_4})+ ...\right) 
\end{equation}
where $\mathcal{L}_M^{kin}$ is the kinematic part of lagrangian of matter filed and "..." includes not only KK modes, but also the interaction between $\psi^{\prime}$ and radion field $R$, which is produced by field redefinition. Since we only consider weakly interacting matter field, we can focus on kinematic term at the leading order. Meanwhile, combining with Equ.(\ref{canR}), we organize Equ.(\ref{cha1}) into a suitable form for Chameleon mechanism:
\begin{equation}
     S_4=\int dx^4 \sqrt{-g_4}\left(\frac{1}{2}M_P^2 R_4 -\frac{1}{2}\partial_u\chi\partial^u\chi-V(\chi) \right) + S_m(\psi^{\prime},\Tilde{g_4})+...
     \label{cha2}
\end{equation}
where  $V(\chi)=2\pi r^2 \Lambda_5 e^{-\sqrt{\frac{2}{3}}\frac{\chi}{M_P}}$ and $\Tilde{g_4}_{\mu \nu}=r e^{-\sqrt{\frac{2}{3}}\frac{\chi}{M_P}}{g_4}_{\mu \nu}$

For non-relativistic matter we have energy density $\Tilde{g_4}^{\mu \nu}\Tilde{T}_{\mu \nu}\sim -\Tilde{\rho}$ where $\Tilde{T}_{\mu \nu}=(2/\sqrt{-\Tilde{g}_4})\delta S_m/ \delta \Tilde{g}_4^{\mu \nu}$ is the physical stress tensor in Jordan frame ($\Tilde{g_4}$). In Einstein frame, we have $\rho = r^2 e^{-2\sqrt{\frac{2}{3}}}\Tilde{\rho}$. However, this energy density is not conserved due to the conservation equation in Einstein frame:
\begin{equation}
    \nabla_{\mu} T^{\mu \nu} = \frac{1}{2}\sqrt{\frac{2}{3}}\frac{1}{M_P}\rho \partial^{\nu}\chi
\end{equation}
The conserved density in Einstein frame is: $\Bar{\rho}=r^{\frac{3}{2}}e^{-\sqrt{\frac{3}{2}}\frac{\chi}{M_P}}\Tilde{\rho}$, which hence is independent of $\chi$. We can then find the equation of motion at the leading order for radion $\chi$ from Equ.(\ref{cha2}):
\begin{equation}
    \partial^2 \chi = \frac{\partial V(\chi)}{\partial \chi} - \sqrt{\frac{1}{6}}\frac{1}{2M_P} \Bar{\rho} r^{\frac{1}{2}}e^{-\sqrt{\frac{1}{6}}\frac{\chi}{M_P}}
\end{equation}

Therefore, we obtained the effective potential:
\begin{equation}
    V_{eff}(\chi)=V(\chi)+r^{\frac{1}{2}}\frac{\Bar{\rho}}{2}e^{-\sqrt{\frac{1}{6}}\frac{\chi}{M_P}}
\end{equation}

Sadly, a negative exponential term shows up, causing the effective potential to continue runaway. If we involve Casimir energy to $V(\chi)$, larger environment density would make the screened mass even lighter. So the Chameleon mechanism can not work for a sole radion field in our model. However, the extra dimension model is friendly to string vacua, so it would be natural for us to include another axion field into our model. Involving axion, we can use the Axio-Chameleon mechanism in Ref.\cite{Brax:2023qyp}, which is a new multi-field screening mechanism appropriate for two light scalar fields (an axion and a Brans-Dicke style dilaton) . The present of axion can safely screen the light dilaton to escape from solar GR test of post-Newtonian deviation. Meanwhile, the mass of axion field can be larger than Neutrino, as shown in Table.1 of Ref.\cite{Brax:2023qyp}. So this axion field will only not change our result on Casimir stabilization of the extra dimension slightly.
\footnote{
Note Axio-Chameleon mechanism requires a non-standard axion-dilaton coupling like $\sim W^2(\chi)(\partial a)^2$. The origin of this coupling might come from target space metric form two scalars $\{ \chi, a \}$, as mentioned in Ref.\cite{Brax:2023qyp}. The detailed mechanism to generate such coupling would beyond the scope of the paper and we'll leave this as future work.}

One the other side, the most stringent requirement is that we need to construct a chiral theory in $5D$, since SM itself is chiral under $SU(2)_{weak}$ gauge symmetry. In this section, we review one well-known way to obtain a chiral SM-like theory as a low energy limit, as is introduced in Ref.\cite{Sundrum:2005jf,Agashe:2006zz, Antoniadis:2023doq}, where we can replace the fifth compact dimension by an interval called "orbifold" of the circle.

The $5D$ chirality problem essentially comes from the representation of $5D$ Clifford algebra for fermions, which is given as:
\begin{equation}
    \{\Gamma_{M},\Gamma_{N}\}=2\eta_{MN},
\end{equation}
where $\Gamma_{\mu}=\gamma_{\mu}$ and $\Gamma_{5}=-i\gamma_{5}$ are usual $4D$ Dirac matrices. So the $5D$ fermions are necessary 4-component Dirac spinors. The problem is that we've used the parity operator $i\gamma_5$ to fill out the $5D$ Clifford algebra so there are no other parity operators left analogous to $\gamma_5$ (parity operator is unique in even dimensional gamma matrices). So we can't find a $5D$ helicity projection operator similar to what we did in $4D$ ($P_{R/L}=\frac{1}{2}(1\pm \gamma_{5})$). Therefore, one can never construct a chiral theory using only 4-component Dirac spinors without helicity projection operators. This is the chirality problem in $5D$.

One way out is to compactify our theory on an orbifold instead of a circle, where we impose an extra $\mathbb{Z}_2$ symmetry on the compactified dimension. To see how orbifold helps, let's write down the kinetic term for fermions:
\begin{equation}
    \mathcal{L}_{kine} \supset i\bar{\psi}_R \partial_{5}\psi_L + i\bar{\psi}_L \partial_{5}\psi_R,
\end{equation}
where we've ignored an overall factor of $i$. Since $\partial_5$ is odd under $\mathbb{Z}_2$, we need one of $\psi_R$ and $\psi_L$ to be odd and another one to be even under $\mathbb{Z}_2$, so that the lagrangian is $\mathbb{Z}_2$ invariant. Therefore, let's decompose $\psi_R$ and $\psi_L$ as:
\begin{equation}
    \begin{aligned}
      \psi_R (x^{\mu},x^5)=\sum_n \psi_R^n(x^{\mu}) sin(n x^5), \\
      \psi_L (x^{\mu},x^5)=\sum_n \psi_L^n(x^{\mu}) cos(n x^5).
    \end{aligned}
\end{equation}
Then in the low energy limit, we can obtain a non-vanishing zero mode for $\psi_L$, while $\psi_R=0$ at $n=0$. 

 A mass term like $\bar{\psi}_R \psi_L + h.c.$ would necessarily breaks $\mathbb{Z}_2$ symmetry. To obtain a mass-like term, we can couple fermions with a pseudo-scalar ($\mathbb{Z}_2$ odd in $\phi$ dimension). The lagrangian then goes like:
\begin{equation}
    \mathcal{L} \supset y_a a \bar{\psi}_R \psi_L + y_a a \bar{\psi}_L \psi_R + V(a)
\end{equation}
where $y_a$ is the coupling constant, $a$ is the pseudo-scalar field and $V(a)$ is the potential for it. If the potential $V(a)$ forms a vacuum expectation value in $\phi$ dimension at $a= \langle a \rangle \neq 0$, it would spontaneously breaks $\mathbb{Z}_2$ symmetry and thus gives mass to fermions proportional to $\langle a \rangle$. Note that since scalar field $a$ is $\mathbb{Z}_2$ odd, so does  $\langle a \rangle$. Thus we at last added a $\mathbb{Z}_2$ odd mass term for fermions. For detail discussion on KK decomposition on massive fermions case, please check Ref.\cite{Agashe:2006zz,Georgi:2000wb}. Meanwhile, Ref.\cite{Scrucca:2003ra} gives many other tricks one can add to connect our model closer to real world.

Finally, we note that adding this extra pseudo-scalar field will not change our previous result dramatically. We only add one extra boson d.o.f., so $n_B$ in equ.(\ref{radpoten}) is just changed from 8 to 9, which is a minor modification to our final result.

\section{Summary}

In summary, we have discussed a $4D$ effective theory in $5D$ SM+GR compactified on a circle of radius R. The effective $4D$ cosmological constant might make the classical potential runaway and decompactify the circle. To stabilize the compact dimension, we considered non-trivial 1-loop vacuum Casimir energy from fields wrapping around the compact direction. The contribution to such Casimir energy comes for three sides: a) massless bosonic fields: graviton + $U(1)$ gauge field with d.o.f. $n_B=8$ in total.  b) fermionic fields with mass less than $1/R_c$: 3 Dirac neutrinos with d.o.f. $n_F=4$ for each neutrino. c) $5D$ cosmological constant. 

In Sec.\ref{result},  we discovered a $dS_4 \times S^1$ vacuum  at $R_c \approx 16.7 eV^{-1}$ by taking the normal hierarchy spectrum for neutrinos (with one massless neutrino) and setting $\Lambda_5 = 3\times 10^{-12}eV^5$ (see Fig.\ref{potential-fig}). To recover the observed value of $4D$ cosmological constant in our universe, we set $r^2 \approx 107 eV^{-2}$. As a $5D$ cosmological constant is not well motivated by string theory, we also considered a rolling scalar field in replacement of $5D$ cosmological constant. The result is almost the same as before. In Sec.\ref{chirality}, we briefly comment on two problems when connecting our model to real $4D$ SM+GR. Firstly, the mass of our radion field is below Hubble scale and would violate the solar system tests of GR. We believe some screening mechanisms are needed. For example, Chameleon mechanism might suitable for our radion field. Secondly, $5D$ spacetime has chirality problem. We  mentioned a possible mechanism to recover chirality from $5D$ in low energy limit, which requires us to compactify our theory on an orbifold and might need an extra pseudo-scalar field to obtain a mass term for fermions. \\

Several comments and discussions along the future prospects are listed:
\begin{itemize}
\item
It would be intriguing to add more degrees of freedom into our $4D$ effective theory. At smaller size of radius, heavier SM particles will contribute to Casimir energy. For instance, electron contribution becomes important at $R \sim 1/m_e$. At $R\sim \Lambda_{QCD}^{-1}$, QCD mesons will come into play and a non-perturbative analysis is needed. Meanwhile, we can also explore new degrees of freedom from Beyond Standard Model physics, such as axions and supersymmetric particles, see \cite{Arkani-Hamed:2007ryu, Anchordoqui:2023wkm, Falkowski:2000er,Falkowski:2000yq, Falkowski:2001sq}.

\item 
It will be interesting to discuss the application of this effective potential in the cosmology \cite{Anchordoqui:2023tln,Ning:2023ybc} and in the context of holography\cite{Conlon:2021cjk, Apers:2022tfm,Ning:2022zqx,Ning:2024apa}, for a nice and comprehensive review see \cite{Cicoli:2023opf}. It might be also intriguing to higher dimensional generalization \cite{Burgess:2023pnk} and WdW perspective\cite{Blacker:2023ezy, Blacker:2024rje, Blacker:2023oan}. 

\item 
Detailed implement of screen mechanisms to the radion field would be interesting. We think Chameleon mechanism might be the easiest one to start with, since it only requires a non-trivial coupling between scalar field and the environment matter 

\item 
The origin of a non-trivial pseudo-scalar potential that breaks $\mathbb{Z}_2$ symmetry on the orbifold is worth discussing. For example, if such scalar field have a slow-roll potential that results in inflation, such potential will also spontaneously breaks $\mathbb{Z}_2$ symmetry.

\end{itemize} 

\acknowledgments

We would like to specially thank Fernando Quevedo for his warm encouragements and many valuable feedbacks and suggestions on an early draft of
this paper. It is also a pleasure to thank Joseph Conlon for useful comments on the draft and Georges Obied for some initial collaborations. Finally we want to thank Hong-Jian He and Frank Petriello for useful conversations and feedbacks. C.C. would like to acknowledge the generous support of the Cambridge University Department of Applied Mathematics and Theoretical Physics and Fitzwilliam College, and Northwestern University Department of Physics and Astronomy and Weinberg College for Arts and Sciences. S.N wants to thank the constant supports from Oxford Physics Department and New College. 


\begin{appendix}
\section{Casimir Energy}
\label{appen}

In this appendix, we derived the 1-loop Casimir energy formula contributed to energy momentum tensor following Ref.\cite{Arkani-Hamed:2007ryu}. We start with a free scalar field with mass $m$ in $d$ dimensional spacetime. The lagrangian reads:
\begin{equation}
    L_{M}=-\frac{1}{2}\partial_\mu\phi\partial^\mu\phi-\frac{1}{2}m^2\phi^2.
\end{equation}
The expectation value of energy momentum tensor takes the form:
\begin{equation}
\begin{aligned}
    \langle T_{\mu\nu}\rangle&=\langle L_M g_{\mu\nu}-2\frac{\delta L_M}{\delta g_{\mu\nu}}\rangle\\
    &=\lim_{x'\rightarrow x}\left(\frac{1}{2}(\partial_\mu\partial_\nu'+\partial_\nu\partial_\mu')-\frac{1}{2}g_{\mu\nu}(\partial^\rho\partial_\rho'+m^2)\right)G(x-x')\label{1},\\
    \end{aligned}
\end{equation}
where $G(x-x')=\langle \phi(x)\phi(x')\rangle$ is the free field propagator. Note that by symmetry $ G(x-x')=G(|x-x'|)$, where  $|x-x'|$ is the distance between two points. When one dimension is compact (call $x^5=\phi$ the compact dimension), we can get Casimir energy from field with different winding number on $\hat{\phi}$ direction:
\begin{equation}
    G(x-x')=\sum_n G_{n}(x-x'+2\pi Rn \hat{\phi}),
\end{equation}
where $n$ ranges from $-\infty$ to $\infty$ without 0, since $R$-independent Casimir contribution should be absorbed into $5D$ cosmological constant.
Note that for massive scalar field:
\begin{equation}
    G(x-x')=\int \frac{d^{d}k}{(2\pi)^d}\frac{e^{ik(x-x')}}{k^2+m^2},
    \label{Green}
\end{equation}
then the second term in \ref{1} goes like:
\begin{equation}
    \begin{aligned}
        &=\frac{1}{2}g_{\mu\nu}(\partial^\rho\partial_\rho'+m^2)\int \frac{d^{d}k}{(2\pi)^d}\frac{e^{ik(x-x')}}{k^2+m^2}\\
        &=\frac{1}{2}g_{\mu\nu}\int \frac{d^{d}k}{(2\pi)^d}e^{ik(x-x')}\\
        &=\frac{1}{2}g_{\mu\nu}\delta^{d-1}(x^{\alpha}-x'^{\alpha})\delta(\phi-\phi^{'}),\\
    \end{aligned}
\end{equation}
where $\alpha$ ranges from 1 to $d-1$. For $G(x-x')=G_{n}(x-x'+2\pi Rn \hat{\phi})$, we have:
\begin{equation}
    \begin{aligned}
        &\frac{1}{2}g_{\mu\nu}(\partial^\rho\partial_\rho'+m^2)G_n(x-x'+2\pi Rn \hat{y})\\
        &=\frac{1}{2}g_{\mu\nu}\delta^{d-1}(x^{\alpha}-x'^{\alpha})\delta(\phi-\phi^{'}+2\pi Rn),\\
    \end{aligned}
\end{equation}
which vanishes in the limit of $x\rightarrow x'$ because $\delta(2\pi Rn)$ is only nonzero when $n$ goes to 0.
Then we calculate the energy momentum:
\begin{equation}
    \begin{aligned}
        \langle T_{\mu\nu}\rangle &=\lim_{x'\rightarrow x}\left(\frac{1}{2}(\partial_\mu\partial_\nu'+\partial_\nu\partial_\mu')-\frac{1}{2}g_{\mu\nu}(\partial^\rho\partial_\rho'+m^2)\right)\sum_n G_n(x-x'+2\pi Rn\hat{\phi})\\
    &=\sum_n\partial_\mu\partial_\nu G_n(x-x'+2\pi Rn\hat{\phi})|_{x=x'}.\\
    \end{aligned}
\end{equation}
For $\mu,\nu=1,...,d-1$, we have:
\begin{equation}
    \begin{aligned}
        &\partial_\mu\partial_\nu G_n(x-x'+2\pi Rn\hat{\phi})|_{x=x'}\\
        &= \partial_\mu\partial_\nu G_n(|x-x'+2\pi Rn\hat{\phi}|)|_{x=x'}\\
        &=2\partial_\mu(\frac{\partial G_n((x-x')^2+\phi_n^2)}{\partial r^2}x_\nu)\Big{|}_{x=x',\phi_n=2\pi Rn\hat{\phi}}\\
        &=2\frac{\partial G_n((x-x')^2+\phi_n^2)}{\partial r^2}\eta_{\mu\nu}+4\frac{\partial^2 G_n((x-x')^2+\phi_n^2)}{\partial^2 r^2}(x-x')_\mu(x-x')_\nu \Big{|}_{x=x',\phi_n=2\pi Rn\hat{\phi}}\\
        &= 2\frac{\partial G_n(\phi_n^2)}{\partial \phi_n^2}\eta_{\mu\nu}\Big{|}_{\phi_n=2\pi Rn\hat{\phi}}\\
        &=\rho_n(R)\eta_{\mu\nu},
    \end{aligned}
\end{equation}
where we defined the Casimir density:
\begin{equation}
\rho_n(R):= 2\frac{\partial G_n(\phi_n^2)}{\partial \phi_n^2} \Big{|}_{\phi_n=2\pi Rn\hat{\phi}}
\end{equation}
For $\mu=1,2,...,d-1, \nu=d$, we have:
\begin{equation}
\begin{aligned}
        &\partial_{\mu}\partial_d G(x-x'+2\pi Rn\hat{\phi})|_{x=x'}\\
        &= \partial_\mu\partial_d G(|x-x'+2\pi Rn\hat{\phi}|)|_{x=x'}\\
        &=2\partial_\mu(\frac{\partial G((x-x')^2+\phi_n^2)}{\partial r^2}\phi_n) \Big{|}_{x=x',\phi_n=2\pi Rn\hat{\phi}}\\
        &=2\frac{\partial G((x-x')^2+\phi_n^2)}{\partial r^2}\eta_{\mu d}+4\frac{\partial^2 G((x-x')^2+\phi_n^2)}{\partial^2 r^2}(x-x')_\mu\phi_n \Big{|}_{x=x',\phi_n=2\pi Rn\hat{\phi}}\\
        &= 0
    \end{aligned}
\end{equation}
For $\mu=d, \nu=d$, we have:
\begin{equation}
    \begin{aligned}
        &\partial_d\partial_d G_n(x-x'+2\pi Rn\hat{\phi})|_{x=x'}\\
        &= \partial_d\partial_d G_n(|x-x'+2\pi Rn\hat{\phi}|)|_{x=x'}\\
        &=2\partial_d(\frac{\partial G_n((x-x')^2+\phi_n^2)}{\partial r^2}\phi_n)\Big{|}_{x=x',\phi_n=2\pi Rn\hat{\phi}}\\
        &=2\frac{\partial G_n((x-x')^2+\phi_n^2)}{\partial r^2}\eta_{\mu d}+4\frac{\partial^2 G_n((x-x')^2+\phi_n^2)}{\partial^2 r^2}\phi_n^2 \Big{|}_{x=x',\phi_n=2\pi Rn\hat{\phi}}\\
        &=2\frac{\partial G_n((x-x')^2+\phi_n^2)}{\partial r^2}\eta_{\mu d}+2\frac{\partial}{\partial R}\frac{\partial G_n((x-x')^2+\phi_n^2)}{\partial r^2}R \Big{|}_{x=x',\phi_n=2\pi Rn\hat{\phi}}\\
        &=2\frac{\partial G_n(\phi_n^2)}{\partial \phi_n^2}\eta_{\mu\nu}|_{x=x',\phi_n=2\pi Rn\hat{\phi}}+R\frac{\partial}{\partial R}2\frac{\partial G_n(\phi_n^2)}{\partial \phi_n^2}\eta_{\mu\nu}|_{x=x',\phi_n=2\pi Rn\hat{\phi}}\\
        &= \rho_n(R)+R\frac{\partial \rho_n(R)}{\partial R}\\
    \end{aligned}
\end{equation}
where we used the formula:
\begin{equation}
    \frac{\partial}{\partial r^2}=\frac{\partial }{\partial R}\frac{\partial R}{\partial r^2}=\frac{\partial}{\partial R}\frac{R}{2y_n^2}
\end{equation}
Therefore, the formula of the Casimir energy momentum tensor can be cast into the form:
\begin{equation}
\begin{aligned}
    T_{\mu\nu}&=-\sum_{n} \left(\rho_n(R)\eta_{\mu\nu}+R\rho_n'(R)\delta_\mu^\phi\delta_\nu^\phi \right)\\
    \end{aligned}
\end{equation}

The green function for $D$ dimensional scalar field is given by \ref{Green}. When it is massless, the green functions reads:
\begin{equation}
    G_{D}(r^2)=\frac{\Gamma(\frac{D}{2}-1)}{4\pi^{\frac{D}{2}}}\frac{1}{r^{D-2}}
\end{equation}
For D=5, we have:
\begin{equation}
    G_{5}(\phi_n^2)=\frac{\Gamma(\frac{5}{2}-1)}{4\pi^{\frac{5}{2}}}\frac{1}{\phi_n^{3}}=\frac{1}{8\pi^{2}}\frac{1}{\phi_n^{3}}\Big{|}_{\phi_n=2\pi nR}=\frac{1}{8\pi^2}\frac{1}{(2\pi n R)^3},
\end{equation}
Hence the Casimir energy density for massless scalar field takes the form:
\begin{equation}
    \rho_n(R)=-\frac{2}{8\pi^2}\cdot\frac{3}{2}\sum_{n}\frac{1}{(2\pi Rn)^5}=-\frac{3}{4\pi^2}\zeta(5)\frac{1}{(2\pi R)^5} 
\end{equation}
where $\zeta(5)=1.036927..$ is the Riemann zeta function at $z=5$. 

For massive scalar field, the green function in $D$ dimension goes like:
\begin{equation}
    G_D(m,r^2)=\frac{m^(D-2)}{(2\pi)^{\frac{D}{2}}}\frac{K_{\frac{d}{2}-1}(mr)}{(mr)^{\frac{d}{2}-1}}
\end{equation}
where m is the mass of the scalar field and $K_{\nu}(z)$ is the modified bessel function of the second kind defined in Equ.(\ref{Besselk}). The Casimir energy density for massive scalar at $D=5$ now reads:
\begin{equation}
    \rho_n(m,R)=-\frac{2m^5}{(2\pi)^{\frac{5}{2}}}\frac{K_{\frac{5}{2}}(2\pi Rmn)}{(2\pi Rmn)^{\frac{5}{2}}}
\end{equation}

\end{appendix}

\bibliographystyle{JHEP}
\bibliography{bibliography}

@article{Arkani-Hamed:2007ryu,
    author = "Arkani-Hamed, Nima and Dubovsky, Sergei and Nicolis, Alberto and Villadoro, Giovanni",
    title = "{Quantum Horizons of the Standard Model Landscape}",
    eprint = "hep-th/0703067",
    archivePrefix = "arXiv",
    doi = "10.1088/1126-6708/2007/06/078",
    journal = "JHEP",
    volume = "06",
    pages = "078",
    year = "2007"
}

@article{Randall:1999ee,
    author = "Randall, Lisa and Sundrum, Raman",
    title = "{A Large mass hierarchy from a small extra dimension}",
    eprint = "hep-ph/9905221",
    archivePrefix = "arXiv",
    reportNumber = "MIT-CTP-2860, PUPT-1860, BUHEP-99-9",
    doi = "10.1103/PhysRevLett.83.3370",
    journal = "Phys. Rev. Lett.",
    volume = "83",
    pages = "3370--3373",
    year = "1999"
}

@article{Montero:2022prj,
    author = "Montero, Miguel and Vafa, Cumrun and Valenzuela, Irene",
    title = "{The dark dimension and the Swampland}",
    eprint = "2205.12293",
    archivePrefix = "arXiv",
    primaryClass = "hep-th",
    doi = "10.1007/JHEP02(2023)022",
    journal = "JHEP",
    volume = "02",
    pages = "022",
    year = "2023"
}

@article{Ooguri:2006in,
    author = "Ooguri, Hirosi and Vafa, Cumrun",
    title = "{On the Geometry of the String Landscape and the Swampland}",
    eprint = "hep-th/0605264",
    archivePrefix = "arXiv",
    reportNumber = "CALT-68-2600, HUTP-06-A017",
    doi = "10.1016/j.nuclphysb.2006.10.033",
    journal = "Nucl. Phys. B",
    volume = "766",
    pages = "21--33",
    year = "2007"
}

@article{Quevedo:2002xw,
    author = "Quevedo, Fernando",
    editor = "de Wit, B. and Vandoren, S.",
    title = "{Lectures on string/brane cosmology}",
    eprint = "hep-th/0210292",
    archivePrefix = "arXiv",
    reportNumber = "DAMTP-2002-128",
    doi = "10.1088/0264-9381/19/22/304",
    journal = "Class. Quant. Grav.",
    volume = "19",
    pages = "5721--5779",
    year = "2002"
}

@article{Quevedo:1997uy,
    author = "Quevedo, Fernando",
    editor = "Cvetic, M. and Langacker, P.",
    title = "{Superstring phenomenology: An Overview}",
    eprint = "hep-ph/9707434",
    archivePrefix = "arXiv",
    reportNumber = "IFUNAM-FT97-11",
    doi = "10.1016/S0920-5632(97)00650-6",
    journal = "Nucl. Phys. B Proc. Suppl.",
    volume = "62",
    pages = "134--143",
    year = "1998"
}

@article{Quevedo:1996sv,
    author = "Quevedo, Fernando",
    editor = "D'Olivo, J. C. and Fernandez, A. and Perez, M. A.",
    title = "{Lectures on superstring phenomenology}",
    eprint = "hep-th/9603074",
    archivePrefix = "arXiv",
    reportNumber = "CERN-TH-96-65",
    doi = "10.1063/1.49735",
    journal = "AIP Conf. Proc.",
    volume = "359",
    pages = "202--242",
    year = "1996"
}

@article{Burgess:2023pnk,
    author = "Burgess, C. P. and Quevedo, F.",
    title = "{Perils of Towers in the Swamp: Dark Dimensions and the Robustness of Effective Field Theories}",
    eprint = "2304.03902",
    archivePrefix = "arXiv",
    primaryClass = "hep-th",
    month = "4",
    year = "2023"
}

@article{Anchordoqui:2023wkm,
    author = "Anchordoqui, Luis A. and Antoniadis, Ignatios and Cunat, Jules",
    title = "{The Dark Dimension and the Standard Model Landscape}",
    eprint = "2306.16491",
    archivePrefix = "arXiv",
    primaryClass = "hep-ph",
    month = "6",
    year = "2023"
}

@article{ParticleDataGroup:2022pth,
    author = "Workman, R. L. and others",
    collaboration = "Particle Data Group",
    title = "{Review of Particle Physics}",
    doi = "10.1093/ptep/ptac097",
    journal = "PTEP",
    volume = "2022",
    pages = "083C01",
    year = "2022"
}

@article{Vainshtein:1972sx,
    author = "Vainshtein, A. I.",
    title = "{To the problem of nonvanishing gravitation mass}",
    doi = "10.1016/0370-2693(72)90147-5",
    journal = "Phys. Lett. B",
    volume = "39",
    pages = "393--394",
    year = "1972"
}

@article{Khoury:2003aq,
    author = "Khoury, Justin and Weltman, Amanda",
    title = "{Chameleon fields: Awaiting surprises for tests of gravity in space}",
    eprint = "astro-ph/0309300",
    archivePrefix = "arXiv",
    doi = "10.1103/PhysRevLett.93.171104",
    journal = "Phys. Rev. Lett.",
    volume = "93",
    pages = "171104",
    year = "2004"
}

@article{Khoury:2003rn,
    author = "Khoury, Justin and Weltman, Amanda",
    title = "{Chameleon cosmology}",
    eprint = "astro-ph/0309411",
    archivePrefix = "arXiv",
    doi = "10.1103/PhysRevD.69.044026",
    journal = "Phys. Rev. D",
    volume = "69",
    pages = "044026",
    year = "2004"
}

@article{Cicoli:2023opf,
    author = "Cicoli, Michele and Conlon, Joseph P. and Maharana, Anshuman and Parameswaran, Susha and Quevedo, Fernando and Zavala, Ivonne",
    title = "{String Cosmology: from the Early Universe to Today}",
    eprint = "2303.04819",
    archivePrefix = "arXiv",
    primaryClass = "hep-th",
    month = "3",
    year = "2023"
}

@article{Anchordoqui:2023tln,
    author = "Anchordoqui, Luis A. and Antoniadis, Ignatios and Lust, Dieter",
    title = "{Fuzzy Dark Matter, the Dark Dimension, and the Pulsar Timing Array Signal}",
    eprint = "2307.01100",
    archivePrefix = "arXiv",
    primaryClass = "hep-ph",
    reportNumber = "MPP-2023-139, LMU-ASC 24/23",
    month = "7",
    year = "2023"
}

@article{Hinterbichler:2010es,
    author = "Hinterbichler, Kurt and Khoury, Justin",
    title = "{Symmetron Fields: Screening Long-Range Forces Through Local Symmetry Restoration}",
    eprint = "1001.4525",
    archivePrefix = "arXiv",
    primaryClass = "hep-th",
    doi = "10.1103/PhysRevLett.104.231301",
    journal = "Phys. Rev. Lett.",
    volume = "104",
    pages = "231301",
    year = "2010"
}

@article{Burgess:2021qti,
    author = "Burgess, C. P. and Quevedo, F.",
    title = "{Axion homeopathy: screening dilaton interactions}",
    eprint = "2110.10352",
    archivePrefix = "arXiv",
    primaryClass = "hep-th",
    reportNumber = "CERN-TH-2021-176",
    doi = "10.1088/1475-7516/2022/04/007",
    journal = "JCAP",
    volume = "04",
    number = "04",
    pages = "007",
    year = "2022"
}

@inproceedings{Sundrum:2005jf,
    author = "Sundrum, Raman",
    title = "{Tasi 2004 lectures: To the fifth dimension and back}",
    booktitle = "{Theoretical Advanced Study Institute in Elementary Particle Physics}: {Physics in D $\geqq$ 4}",
    eprint = "hep-th/0508134",
    archivePrefix = "arXiv",
    pages = "585--630",
    month = "8",
    year = "2005"
}

@inproceedings{Agashe:2006zz,
    author = "Agashe, Kaustubh",
    title = "{Extra dimensions}",
    booktitle = "{Theoretical Advanced Study Institute in Elementary Particle Physics}: {Exploring New Frontiers Using Colliders and Neutrinos}",
    pages = "1--48",
    month = "6",
    year = "2006"
}

@article{Georgi:2000wb,
    author = "Georgi, Howard and Grant, Aaron K. and Hailu, Girma",
    title = "{Chiral fermions, orbifolds, scalars and fat branes}",
    eprint = "hep-ph/0007350",
    archivePrefix = "arXiv",
    reportNumber = "HUTP-00-A029",
    doi = "10.1103/PhysRevD.63.064027",
    journal = "Phys. Rev. D",
    volume = "63",
    pages = "064027",
    year = "2001"
}

@article{Scrucca:2003ra,
    author = "Scrucca, Claudio A. and Serone, Marco and Silvestrini, Luca",
    title = "{Electroweak symmetry breaking and fermion masses from extra dimensions}",
    eprint = "hep-ph/0304220",
    archivePrefix = "arXiv",
    reportNumber = "CERN-TH-2003-028, ROMA-1354-03, SISSA-32-2003-EP",
    doi = "10.1016/j.nuclphysb.2003.07.013",
    journal = "Nucl. Phys. B",
    volume = "669",
    pages = "128--158",
    year = "2003"
}

@inproceedings{Shaposhnikov:2007nj,
    author = "Shaposhnikov, Mikhail",
    title = "{Is there a new physics between electroweak and Planck scales?}",
    booktitle = "{Astroparticle Physics: Current Issues, 2007 (APCI07)}",
    eprint = "0708.3550",
    archivePrefix = "arXiv",
    primaryClass = "hep-th",
    month = "8",
    year = "2007"
}

@article{Nielsen:2012pu,
    author = "Nielsen, Holger Bech",
    editor = "Mankoc Borstnik, Norma Susana and Khlopov, M. Y. and Nielsen, Holger Bech and Lukman, Dragan",
    title = "{PREdicted the Higgs Mass}",
    eprint = "1212.5716",
    archivePrefix = "arXiv",
    primaryClass = "hep-ph",
    journal = "Bled Workshops Phys.",
    volume = "13",
    number = "2",
    pages = "94--126",
    year = "2012"
}

@article{Masina:2012tz,
    author = "Masina, Isabella",
    title = "{Higgs boson and top quark masses as tests of electroweak vacuum stability}",
    eprint = "1209.0393",
    archivePrefix = "arXiv",
    primaryClass = "hep-ph",
    reportNumber = "CP3-Origins-2012-22, DIAS-2012-23",
    doi = "10.1103/PhysRevD.87.053001",
    journal = "Phys. Rev. D",
    volume = "87",
    number = "5",
    pages = "053001",
    year = "2013"
}

@article{Degrassi:2012ry,
    author = "Degrassi, Giuseppe and Di Vita, Stefano and Elias-Miro, Joan and Espinosa, Jose R. and Giudice, Gian F. and Isidori, Gino and Strumia, Alessandro",
    title = "{Higgs mass and vacuum stability in the Standard Model at NNLO}",
    eprint = "1205.6497",
    archivePrefix = "arXiv",
    primaryClass = "hep-ph",
    reportNumber = "CERN-PH-TH-2012-134, RM3-TH-12-9",
    doi = "10.1007/JHEP08(2012)098",
    journal = "JHEP",
    volume = "08",
    pages = "098",
    year = "2012"
}

@article{Peccei:1977ur,
    author = "Peccei, R. D. and Quinn, Helen R.",
    title = "{Constraints Imposed by CP Conservation in the Presence of Instantons}",
    reportNumber = "ITP-572-STANFORD",
    doi = "10.1103/PhysRevD.16.1791",
    journal = "Phys. Rev. D",
    volume = "16",
    pages = "1791--1797",
    year = "1977"
}

@article{Shifman:1979if,
    author = "Shifman, Mikhail A. and Vainshtein, A. I. and Zakharov, Valentin I.",
    title = "{Can Confinement Ensure Natural CP Invariance of Strong Interactions?}",
    reportNumber = "ITEP-64-1979",
    doi = "10.1016/0550-3213(80)90209-6",
    journal = "Nucl. Phys. B",
    volume = "166",
    pages = "493--506",
    year = "1980"
}

@article{Kim:2008hd,
    author = "Kim, Jihn E. and Carosi, Gianpaolo",
    title = "{Axions and the Strong CP Problem}",
    eprint = "0807.3125",
    archivePrefix = "arXiv",
    primaryClass = "hep-ph",
    doi = "10.1103/RevModPhys.82.557",
    journal = "Rev. Mod. Phys.",
    volume = "82",
    pages = "557--602",
    year = "2010",
    note = "[Erratum: Rev.Mod.Phys. 91, 049902 (2019)]"
}

@article{tHooft:1979rat,
    author = "'t Hooft, Gerard",
    editor = "'t Hooft, Gerard and Itzykson, C. and Jaffe, A. and Lehmann, H. and Mitter, P. K. and Singer, I. M. and Stora, R.",
    title = "{Naturalness, chiral symmetry, and spontaneous chiral symmetry breaking}",
    reportNumber = "PRINT-80-0083 (UTRECHT)",
    doi = "10.1007/978-1-4684-7571-5_9",
    journal = "NATO Sci. Ser. B",
    volume = "59",
    pages = "135--157",
    year = "1980"
}

@article{Antoniadis:2023doq,
    author = "Antoniadis, Ignatios and Benakli, Karim",
    title = "{Extra Dimensions and Physics of Low Scale Strings}",
    eprint = "2305.11604",
    archivePrefix = "arXiv",
    primaryClass = "hep-ph",
    month = "5",
    year = "2023"
}

@article{Brax:2023qyp,
    author = "Brax, Philippe and Burgess, C. P. and Quevedo, F.",
    title = "{Axio-Chameleons: A Novel String-Friendly Multi-field Screening Mechanism}",
    eprint = "2310.02092",
    archivePrefix = "arXiv",
    primaryClass = "hep-th",
    month = "10",
    year = "2023"
}

@article{Branchina:2023ogv,
    author = "Branchina, Carlo and Branchina, Vincenzo and Contino, Filippo and Pernace, Arcangelo",
    title = "{Does the Cosmological Constant really indicate the existence of a Dark Dimension?}",
    eprint = "2308.16548",
    archivePrefix = "arXiv",
    primaryClass = "hep-th",
    month = "8",
    year = "2023"
}

@article{Antoniadis:1998ig,
    author = "Antoniadis, Ignatios and Arkani-Hamed, Nima and Dimopoulos, Savas and Dvali, G. R.",
    title = "{New dimensions at a millimeter to a Fermi and superstrings at a TeV}",
    eprint = "hep-ph/9804398",
    archivePrefix = "arXiv",
    reportNumber = "SLAC-PUB-7801, SU-ITP-98-28, CPTH-S608-0498, IC-98-39",
    doi = "10.1016/S0370-2693(98)00860-0",
    journal = "Phys. Lett. B",
    volume = "436",
    pages = "257--263",
    year = "1998"
}

@article{Arkani-Hamed:1998wuz,
    author = "Arkani-Hamed, Nima and Dimopoulos, Savas and Dvali, G. R. and March-Russell, John",
    title = "{Neutrino masses from large extra dimensions}",
    eprint = "hep-ph/9811448",
    archivePrefix = "arXiv",
    reportNumber = "SLAC-PUB-8014, SU-ITP-98-64",
    doi = "10.1103/PhysRevD.65.024032",
    journal = "Phys. Rev. D",
    volume = "65",
    pages = "024032",
    year = "2001"
}

@article{Randall:1998uk,
    author = "Randall, Lisa and Sundrum, Raman",
    title = "{Out of this world supersymmetry breaking}",
    eprint = "hep-th/9810155",
    archivePrefix = "arXiv",
    reportNumber = "MIT-CTP-2788, PUPT-1815, BUHEP-98-26",
    doi = "10.1016/S0550-3213(99)00359-4",
    journal = "Nucl. Phys. B",
    volume = "557",
    pages = "79--118",
    year = "1999"
}

@article{Falkowski:2000er,
    author = "Falkowski, Adam and Lalak, Zygmunt and Pokorski, Stefan",
    title = "{Supersymmetrizing branes with bulk in five-dimensional supergravity}",
    eprint = "hep-th/0004093",
    archivePrefix = "arXiv",
    doi = "10.1016/S0370-2693(00)00995-3",
    journal = "Phys. Lett. B",
    volume = "491",
    pages = "172--182",
    year = "2000"
}

@article{Falkowski:2000yq,
    author = "Falkowski, Adam and Lalak, Zygmunt and Pokorski, Stefan",
    title = "{Five-dimensional gauged supergravities with universal hypermultiplet and warped brane worlds}",
    eprint = "hep-th/0009167",
    archivePrefix = "arXiv",
    doi = "10.1016/S0370-2693(01)00269-6",
    journal = "Phys. Lett. B",
    volume = "509",
    pages = "337--345",
    year = "2001"
}

@article{Falkowski:2001sq,
    author = "Falkowski, Adam and Lalak, Zygmunt and Pokorski, Stefan",
    title = "{Four-dimensional supergravities from five-dimensional brane worlds}",
    eprint = "hep-th/0102145",
    archivePrefix = "arXiv",
    reportNumber = "CERN-TH-2000-366",
    doi = "10.1016/S0550-3213(01)00376-5",
    journal = "Nucl. Phys. B",
    volume = "613",
    pages = "189--217",
    year = "2001"
}

@article{Conlon:2021cjk,
    author = "Conlon, Joseph P. and Ning, Sirui and Revello, Filippo",
    title = "{Exploring the holographic Swampland}",
    eprint = "2110.06245",
    archivePrefix = "arXiv",
    primaryClass = "hep-th",
    doi = "10.1007/JHEP04(2022)117",
    journal = "JHEP",
    volume = "04",
    pages = "117",
    year = "2022"
}

@article{Apers:2022tfm,
    author = "Apers, Fien and Conlon, Joseph P. and Ning, Sirui and Revello, Filippo",
    title = "{Integer conformal dimensions for type IIa flux vacua}",
    eprint = "2202.09330",
    archivePrefix = "arXiv",
    primaryClass = "hep-th",
    doi = "10.1103/PhysRevD.105.106029",
    journal = "Phys. Rev. D",
    volume = "105",
    number = "10",
    pages = "106029",
    year = "2022"
}

@article{Ning:2022zqx,
    author = "Ning, Sirui",
    title = "{Holographic perspectives on models of moduli stabilization in M-theory}",
    eprint = "2206.13332",
    archivePrefix = "arXiv",
    primaryClass = "hep-th",
    doi = "10.1007/JHEP09(2022)042",
    journal = "JHEP",
    volume = "09",
    pages = "042",
    year = "2022"
}

@article{Blacker:2023ezy,
    author = "Blacker, Matthew J. and Ning, Sirui",
    title = "{Wheeler DeWitt states of a charged AdS$_{4}$ black hole}",
    eprint = "2308.00040",
    archivePrefix = "arXiv",
    primaryClass = "hep-th",
    doi = "10.1007/JHEP12(2023)002",
    journal = "JHEP",
    volume = "12",
    pages = "002",
    year = "2023"
}

@article{Blacker:2024rje,
    author = "Blacker, Matthew J. and Callebaut, Nele and Hergueta, Blanca and Ning, Sirui",
    title = "{Radial canonical AdS$_3$ gravity and $T\bar{T}$}",
    eprint = "2406.02508",
    archivePrefix = "arXiv",
    primaryClass = "hep-th",
    month = "6",
    year = "2024"
}

@article{Blacker:2023oan,
    author = "Blacker, Matthew J. and Hartnoll, Sean A.",
    title = "{Cosmological quantum states of de Sitter-Schwarzschild are static patch partition functions}",
    eprint = "2304.06865",
    archivePrefix = "arXiv",
    primaryClass = "hep-th",
    doi = "10.1007/JHEP12(2023)025",
    journal = "JHEP",
    volume = "12",
    pages = "025",
    year = "2023"
}

@article{Yang:2018iki,
    author = "Yang, Zhi and Cheng, Long and Hung, Ling-Yan and Ning, Sirui and Bhattacharyya, Arpan",
    title = "{Emergent Lorentz symmetry and the Unruh effect in a Lorentzian fermionic tensor network}",
    eprint = "1805.03071",
    archivePrefix = "arXiv",
    primaryClass = "hep-th",
    reportNumber = "YITP-18-40",
    doi = "10.1103/PhysRevD.99.086007",
    journal = "Phys. Rev. D",
    volume = "99",
    number = "8",
    pages = "086007",
    year = "2019"
}

@misc{BentoMontero2025,
  title        = {An M-theory dS maximum from Casimir energies on Riemann-flat manifolds},
  author       = {Bruno Valeixo Bento and Miguel Montero},
  year         = {2025},
  eprint       = {2507.02037},
  archivePrefix= {arXiv},
  primaryClass = {hep-th},
  note         = {arXiv:2507.02037 [hep-th]}
}

@misc{Rudelius2021,
  title        = {Asymptotic Observables and the Swampland},
  author       = {Tom Rudelius},
  year         = {2021},
  eprint       = {2106.09026},
  archivePrefix= {arXiv},
  primaryClass = {hep-th},
  note         = {arXiv:2106.09026 [hep-th]}
}

@phdthesis{Ning:2024apa,
    author = "Ning, Sirui",
    title = "{String compactification, effective field theory and holography swampland}",
    eprint = "2512.11733",
    archivePrefix = "arXiv",
    primaryClass = "hep-th",
    doi = "10.5287/ora-rvzy0jpee",
    school = "Oxford University, Oxford U.",
    year = "2024"
}

@article{Ning:2023ybc,
    author = "Ning, Sirui and Sou, Chon Man and Wang, Yi",
    title = "{On the decoherence of primordial gravitons}",
    eprint = "2305.08071",
    archivePrefix = "arXiv",
    primaryClass = "hep-th",
    doi = "10.1007/JHEP06(2023)101",
    journal = "JHEP",
    volume = "06",
    pages = "101",
    year = "2023"
}

\end{document}